\documentstyle[12pt] {article}
\title{}
\author{}
\begin{document}
\begin{center}
{\LARGE ASTRO SPACE CENTER}\\
\vspace*{1cm}
{\bf THE TUNNELING, THE SECOND ORDER RELATIVISTIC\\
PHASE TRANSITIONS AND PROBLEM OF THE MACROSCOPIC UNIVERSE ORIGIN}\\
\vspace*{1.0cm}
\hspace*{6.5cm}Preprint ASC-95\\
\hspace*{6.5cm}October, 1995\\
\hspace*{6.5cm}Astro-ph/  \\
\hspace*{6.5cm} Submitted to Intern.J. of Modern Physics D.\\

\vspace*{1.0cm}
{\bf V.V.Burdyuzha ${}^{1}$, Yu.N.Ponomarev ${}^{2}$}\\
\vspace*{.2cm}
{\it Astro Space Centre Lebedev Physical Institute of Russian \\
Academy of Sciences\\
Profsoyuznaya 84/32, 117810 Moscow, Russia}\\
\vspace*{.5cm} 
{\bf O.D.Lalakulich, G.M.Vereshkov}\\  
\vspace*{.2cm}
{\it Scientific Research Institute of Physics\\
Rostov State University\\
Stachki str. 194, 344104, Rostov on Don, Russia}\\
\end{center}
\vspace*{1.0cm}
\baselineskip=6 mm plus 0.1 mm minus 0.1 mm
{\bf ABSTRACT}\\
\\
We propose that the Universe created from "Nothing" with a relatively
small  particles number and it very quick relaxed to quasiequilibrium
state at the Planck parameters. The classic cosmological solution for
this Universe, with the calculation of it ability to be undergo to the
second order relativistic phase transition ( RPT ), has two branches
divided by gap. On one from these branches near to "Nothing" state
the second order RPT isn't possible at GUT scale. Other branch
is thermodynamically instable. The quantum process of tunneling between
the cosmological solution branches  and kinetics of the second order
RPT are investigated by numerical methods. Other quantum
geometrodynamics process (bounce from singularity) taken into consideration 
also. It is shown that
discussed phenomenon with the calculaton of all RPT from scale GUT
($10^{16}$ Gev) to Salam-Weinberg scale ($10^{2}$ Gev) gives the new
cosmological scenarious of the macroscopic Universe origin with 
observable particles number.\\
\\
email: ${}^{1}$ burdyuzh@dpc.asc.rssi.ru \hspace*{.5cm} ${}^{2}$ yupon@dpc.asc.rssi.ru 
\newpage
\end{document}